# Electromagnetic Angular Momentum

Masud Mansuripur, College of Optical Sciences
The University of Arizona, Tucson

**Status**. In addition to energy and linear momentum, electromagnetic (EM) fields can carry spin angular momentum (when circularly or elliptically polarized), and also orbital angular momentum (for example, in the presence of phase vorticity).[1-3] The classical theory of electrodynamics does not distinguish between the two types of angular momentum, treating them on an equal footing.[4-8] In quantum optics, however, while a single photon may possess only one $\hbar$ of spin, it has the capacity to carry $m\hbar$ of orbital angular momentum, $m$ being an arbitrary integer—or, in special cases, a half-integer.[9] As usual, $\hbar$ is Planck's reduced constant.

In the section on EM momentum, we pointed out that the specification of a stress tensor uniquely identifies, in conjunction with Maxwell's macroscopic equations, the densities of EM force and torque as well as those of linear and angular momenta. In the present section, we rely once again on the Einstein-Laub tensor to elucidate certain properties of EM angular momentum, whose density is given by $\mathcal{L}(r,t) = r \times p(r,t)$. Here $p = S(r,t)/c^2$ is the EM linear momentum density, $S = E \times H$ is the Poynting vector, and $c$ is the speed of light in vacuum. Below we discuss two examples of EM angular momentum, one in a static situation, the other in a dynamic context, thus highlighting the universality of the concept.

**Example 1**. Shown in Fig.1(a) is a uniformly-charged hollow sphere of radius $R$ and surface-charge-density $\sigma_s$. The sphere is spinning at a constant angular velocity $\Omega_0$ around the $z$-axis. The total charge of the sphere is $Q = 4\pi R^2 \sigma_s$, while its magnetic dipole moment is $\boldsymbol{m} = (4\pi R^3/3)(\mu_0 \sigma_s R \Omega_0)\hat{z}$. Inside the sphere, the electric and magnetic fields are $E(r) = 0$ and $H(r) = \tfrac{2}{3}(\sigma_s R \Omega_0)\hat{z}$, whereas outside the sphere,

$$E(r) = \sigma_s R^2 \hat{r}/(\varepsilon_0 r^2). \qquad (1)$$

$$H(r) = \tfrac{1}{3}(\sigma_s R^4 \Omega_0)(2\cos\theta\,\hat{r} + \sin\theta\,\hat{\theta})/r^3. \qquad (2)$$

Integrating the angular momentum density $\mathcal{L}(r)$ over the entire space yields the total EM angular momentum of the spinning charged sphere as $\boldsymbol{L} = Q\boldsymbol{m}/(6\pi R)$. For electrons, $Q = -1.6 \times 10^{-19}$ coulomb, $m_z = \mu_0 \mu_B = -4\pi \times 10^{-7} \times 9.27 \times 10^{-24}$ joule·meter/ampere, and $L_z = \tfrac{1}{2}\hbar = 0.527 \times 10^{-34}$ joule·sec. If the electron were a ball of radius $R \cong 1.876$ femtometer, its entire angular momentum would have been electromagnetic.

Next, consider Fig.1(b), where a second uniformly-charged spherical shell of radius $R_0$ carrying a total charge of $-Q$ is placed at the center of the spinning sphere. The $E$-field will now be confined to the region between the two spheres, and the total EM angular momentum of the system is readily found to be

$$\boldsymbol{L} = \tfrac{8\pi}{9}(\mu_0 R^3 \sigma_s^2 \Omega_0)(R^2 - R_0^2)\hat{z}. \qquad (3)$$

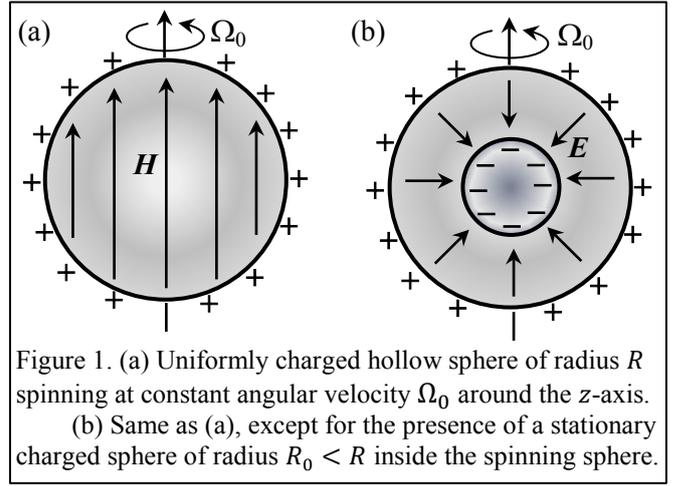

Figure 1. (a) Uniformly charged hollow sphere of radius $R$ spinning at constant angular velocity $\Omega_0$ around the $z$-axis.
(b) Same as (a), except for the presence of a stationary charged sphere of radius $R_0 < R$ inside the spinning sphere.

It is thus seen that, in the limit when $R_0 \to 0$, the two angular momenta, calculated with and without the second sphere, are identical. In one case, the "particle" has a net charge of $Q$ and its EM angular momentum is distributed outside the sphere of radius $R$. In the other case the "particle" (consisting of two spheres) is charge neutral, yet continues to possess a magnetic dipole moment (because the external sphere is spinning). In the latter case, the EM angular momentum is confined to the space between the two spheres.

**Example 2**. Here we consider a single-mode vector spherical harmonic EM wave[10] trapped inside a hollow, perfectly conducting sphere of radius $R$. Choosing the mode's polarization state to be transverse electric (TE) and denoting its frequency by $\omega$, its wavenumber by $k_0 = \omega/c$, and its polar and azimuthal orders by $(\ell, m)$, the field distribution inside the sphere will be

$$\boldsymbol{E}(r,t) = E_0 \frac{J_{\ell+\frac{1}{2}}(k_0 r)}{\sqrt{k_0 r}} \left\{ \frac{P_\ell^m(\cos\theta)}{\sin\theta} \cos(m\varphi - \omega t)\,\hat{\theta} \right.$$
$$\left. + \frac{\sin\theta\, P'^m_\ell(\cos\theta)}{m} \sin(m\varphi - \omega t)\,\hat{\varphi} \right\}. \qquad (4)$$

$$\boldsymbol{H}(r,t) = -\frac{E_0}{Z_0}\left\{ \frac{\ell(\ell+1) J_{\ell+\frac{1}{2}}(k_0 r)}{(k_0 r)^{3/2}} \frac{P_\ell^m(\cos\theta)\cos(m\varphi-\omega t)}{m}\hat{r} \right.$$
$$-\frac{k_0 r J'_{\ell+\frac{1}{2}}(k_0 r) + \tfrac{1}{2} J_{\ell+\frac{1}{2}}(k_0 r)}{(k_0 r)^{3/2}}\left[ \frac{\sin\theta\, P'^m_\ell(\cos\theta)\cos(m\varphi-\omega t)}{m}\hat{\theta} \right.$$
$$\left.\left. + \frac{P_\ell^m(\cos\theta)\sin(m\varphi-\omega t)}{\sin\theta}\hat{\varphi} \right]\right\}. \qquad (5)$$

In the above equations, $E_0$ is the $E$-field amplitude, $E_0/Z_0$ is the $H$-field amplitude, $Z_0 = \sqrt{\mu_0/\varepsilon_0}$ is the impedance of free space, $J_{\ell+\frac{1}{2}}(\rho)$ is a Bessel function of first kind, half-integer order $\ell + \frac{1}{2}$, $P_\ell^m(\zeta)$ is an associated Legendre function of order $(\ell, m)$, and the integers $\ell$ and $1 \le m \le \ell$ are the polar and azimuthal orders of the mode. The primes over $J_{\ell+\frac{1}{2}}(\rho)$ and $P_\ell^m(\zeta)$ indicate differentiation with respect to $\rho$ and $\zeta$. Since the tangential components of $\boldsymbol{E}$ and the perpendicular component of $\boldsymbol{B} = \mu_0 \boldsymbol{H}$ must vanish on the internal surface of the perfect conductor, it is necessary that $k_0 R$ be a zero of $J_{\ell+\frac{1}{2}}(\rho)$.

The field energy-density, $\mathcal{E}(\mathbf{r}, t) = \frac{1}{2}\varepsilon_0 E^2 + \frac{1}{2}\mu_0 H^2$, when integrated over the spherical volume of radius $R$, yields the total energy $\mathcal{E}_{\text{total}}^{(\text{EM})}$ of the mode trapped inside the cavity. In the limit of large $R$ we find

$$\mathcal{E}_{\text{total}}^{(\text{EM})} \approx \left(\frac{RE_0^2}{\mu_0 \omega^2}\right) \frac{\ell(\ell+1)(\ell+m)!}{m^2(\ell+\frac{1}{2})(\ell-m)!}. \quad (6)$$

Similarly, integration of the angular momentum-density $\mathcal{L}(\mathbf{r}, t)$ over the sphere's volume in the limit of large $R$ yields

$$L_z^{(\text{EM})} \approx \left(\frac{RE_0^2}{\mu_0 \omega^3}\right) \frac{\ell(\ell+1)(\ell+m)!}{m(\ell+\frac{1}{2})(\ell-m)!}. \quad (7)$$

Thus, for a sufficiently large sphere, the ratio of orbital angular momentum to EM energy is seen to be

$$L_z^{(\text{EM})} / \mathcal{E}_{\text{total}}^{(\text{EM})} = m/\omega. \quad (8)$$

This is consistent with the quantum-optical picture of photons of energy $\hbar\omega$ and orbital angular momentum $m\hbar\hat{\mathbf{z}}$ being trapped within the cavity.

**Angular Momentum Conservation**. Returning to the system of Fig.1(a), let the sphere's angular velocity be a function of time, $\Omega(t)$, which starts at zero and rises slowly to $\Omega_0$ after a long time. In accordance with Faraday's law $(\nabla \times \mathbf{E} = -\partial \mathbf{B}/\partial t)$, the changing magnetic field inside the sphere induces the $E$-field $\mathbf{E}(\mathbf{r}, t) = -\frac{1}{3}\mu_0 \sigma_s R r \sin\theta\, \Omega'(t)\hat{\boldsymbol{\varphi}}$. The torque acting on the spinning sphere by this $E$-field will then be

$$\mathbf{T}(t) = \int_{\theta=0}^{\pi} \int_{\varphi=0}^{2\pi} \mathbf{r} \times \sigma_s \mathbf{E}(R, \theta, \varphi, t) R^2 \sin\theta\, d\theta d\varphi$$

$$= -\frac{8\pi}{9}\mu_0 R^5 \sigma_s^2 \Omega'(t)\hat{\mathbf{z}}. \quad (9)$$

Integration over time now reveals that the mechanical angular momentum delivered by an external torque during the spin-up process is precisely equal to the EM angular momentum $\mathbf{L}$ residing in the space surrounding the spinning sphere of Fig.1(a). A similar calculation shows that the EM angular momentum in the system of Fig.1(b) is fully accounted for by the mechanical torque acting on the outer (spinning) sphere minus the torque needed to keep the inner sphere stationary.

The above argument, although employed here in conjunction with simple quasi-static systems involving slowly-spinning charged spheres, applies equally well to more complicated dynamic situations.[5,11] In general, any change of the EM angular momentum of a system is always accompanied by a corresponding mechanical torque acting on the material bodies of the system. The existence of this torque ensures that any change in the EM angular momentum is compensated, precisely and instantaneously, by an equal and opposite change in the overall mechanical angular momentum of the system.[12]

**Current and Future Challenges**. To date, essentially all experimental researches on EM angular momentum have been carried out with non-magnetic materials. Since some of the major differences among the various EM stress tensors depend on magnetization, it will be worthwhile to explore the mechanical response of magnetized (or magnetisable) media in the presence of EM fields — in particular, external electric fields, which give rise to the so-called "hidden momentum" in conjunction with the standard Lorentz stress tensor.[13] A topic of fundamental importance is the *distribution* of EM force-density and/or torque-density within material objects, which can only be monitored in soft (i.e., deformable) matter. Although different stress tensors might predict identical behaviour for a rigid object in response to the overall EM force/torque acting on the object, local deformations of soft media are expected to vary depending on the distribution of force and torque densities throughout the object. Precise measurements of this type can thus distinguish among the various stress tensors that have been proposed in the literature.

Another interesting topic is the interaction of light carrying EM angular momentum with rotating bodies such as mirrors, transparent or partially-absorbing plates, and birefringent media.[14,15] This would involve exchanges of angular momentum as well as energy, resulting in a Doppler shift of the emerging photons.

**Advances in Science and Technology to Meet Challenges**. The most dramatic effects of EM linear and angular momenta arise in conjunction with high quality-factor micro-resonators and nano-devices. Advances in micro- and nano-fabrication in tandem with development of novel mechanisms for coupling light and/or microwaves into and out of high-Q cavities will enable sensitive measurements of opto-mechanical interactions. Development of high-quality transparent magnetic materials as well as novel meta-materials with controlled permittivity, permeability, dispersion, birefringence, nonlinear coefficients, etc., will usher in not only a deeper understanding of the fundamentals but also a treasure trove of new applications.

**Concluding Remarks**. While the classical theory of EM angular momentum is well established, there remain opportunities for theoretical contributions to the understanding of nonlinear phenomena, quantum optics of non-paraxial beams, and interactions between light and rotating objects. On the experimental side, one could look forward to precise measurements of EM force-density and torque-density distributions in conjunction with deformable media in order to identify appropriate stress tensors for ordinary matter and also for exotic meta-materials. The bidirectional exchange of angular momentum between EM fields and material objects continues to provide a rich and rewarding platform for exploring novel practical applications.

**References**

[1] L. Allen, S. M. Barnett, and M. J. Padgett, *Optical Angular Momentum*, Institute of Physics Publishing, Bristol, U.K. (2003).

[2] L. Allen, M. W. Beijersbergen, R. J. C. Spreeuw, and J. P. Woerdman, *Phys. Rev. A* **45**, 8185 (1992).

[3] N. B. Simpson, K. Dholakia, L. Allen and M. J. Padgett, *Opt. Lett.* **22**, 52 (1997).

[4] M. Mansuripur, "Angular momentum of circularly polarized light in dielectric media," *Optics Express*, **13**, 5315-24 (2005).

[5] M. Mansuripur, "Electromagnetic force and torque in ponderable media," *Optics Express* **16**, 14821-35 (2008).

[6] S. M. Barnett, "Rotation of electromagnetic fields and the nature of optical angular momentum," *J. Mod. Opt.* **57**, 1339-1343 (2010).

[7] M. Mansuripur, A. R. Zakharian, and E. M. Wright, "Spin and orbital angular momenta of light reflected from a cone," *Phys. Rev. A* **84**, 033813, pp1-12 (2011).

[8] M. Mansuripur, "Spin and orbital angular momenta of electromagnetic waves in free space," *Phys. Rev. A* **84**, 033838, pp1-6 (2011).

[9] M. V. Berry, M. R. Jeffrey, and M. Mansuripur, "Orbital and spin angular momentum in conical diffraction," *J. Opt. A: Pure Appl. Opt.* **7**, 685-690 (2005).

[10] J. D. Jackson, *Classical Electrodynamics*, 3rd edition, Wiley, New York (1999).

[11] M. Mansuripur, "Energy and linear and angular momenta in simple electromagnetic systems," Optical Trapping and Optical Micromanipulation XII, edited by K. Dholakia and G. C. Spalding, *Proc. of SPIE* Vol. **9548**, 95480K ~ 1:24 (2015).

[12] M. Kristensen and J. P. Woerdman, "Is photon angular momentum conserved in a dielectric medium?" *Phys. Rev. Lett.* **72**, 2171-2174 (1994).

[13] M. Mansuripur, "The charge-magnet paradoxes of classical electrodynamics," Spintronics VII, edited by H. J. Drouhin, J. E. Wegrowe, and M. Razeghi, *Proc. of SPIE* Vol. **9167**, 91670J ~ 1:12 (2014).

[14] M. J. Padgett, "The mechanism for energy transfer in the rotational frequency shift of a light beam," *J. Opt. A: Pure Appl. Opt.* **6**, S263-S265 (2004).

[15] M. Mansuripur, "Angular momentum exchange between light and material media deduced from the Doppler shift," Optical Trapping and Optical Micromanipulation IX, edited by K. Dholakia and G. C. Spalding, *Proc. of SPIE* Vol. **8458**, 845805 ~ 1:8 (2012).